# Quanta: a Language for Modeling and Manipulating Information Structures

Bruce Long, Computer Science Department, University of Westminster, London

**Abstract.** We present a theory for modeling the structure of information and a language (Quanta) expressing the theory. Unlike Shannon's information theory, which focuses on the *amount* of information in an information system, we focus on the structure of the information in the system. For example, we can model the information structure corresponding to an algorithm or a physical process such as the structure of a quantum interaction. After a brief discussion of the relation between an evolving state-system and an information structure, we develop an algebra of information pieces (infons) to represent the structure of systems where descriptions of complex systems are constructed from expressions involving descriptions of simpler information systems. We map the theory to the Von Neumann computing model of sequences/conditionals/repetitions, and to the class/object theory of object-oriented programming (OOP).

**Table of Contents**

## 1 Introduction

In the paper *The Structure of Information* [Long] a mathematical model of information is presented. This paper is based upon that theory but is not directly inferred from it. While that paper focuses on representing very simple closed systems, this paper focuses on describing complex open systems such as macro-sized physical or computer objects and systems of objects. Furthermore, while that paper is presented in the language of mathematics, this paper is presented in the language of computer science, i.e. a specification of syntax and semantics. A language (Quanta) is developed for describing objects and classes and an engine is described for using the Quanta descriptions (in documents) to infer what actions or sequences of actions would be identical to some description and to actually *do* those actions to accomplish a task or instantiate a class. While the theory presented here is focused on macro-sized open systems, Quanta can also represent small closed information systems; thus, the name "Quanta."

## 2 The Structure of Information

### *2.1 From States to Bits*

A common method of representing the structure of an information system is to use the formalism of a state machine where states are mapped to states on transitions. The use of group theory to represent state systems is an example [Armstrong]. A reasonable alternative view to a state machine model of state systems is suggested by the fact that a state system, whether in a computer or some "physical" system, can be seen as a container of information. In fact, it is common knowledge in computer science that the number of bits of information (b) is related to the number of states (s) by [Shannon]:

$$s = 2^b.$$

In this paper, a model of the World-as-state-system is given in terms of information structures rather than state correlations[1]. An interesting way to develop such a theory would be to start with a state-transition based formalism such as group theory or the formalism of Petri nets [Murata] and use the above identity to generate a bit-centric view. However, here we shall merely consider, as an example, the operation of computer memory as it evolves deterministically through a sequence of states under the control of a program, and extract a number of "observations" which we shall use to generate the formal theory. The example is not meant to "prove" the observations; rather, it is intended merely to provide an intuitive foundation for questioning the formalism produced.

#### 2.1.1 The Computer Memory Example

In this section, we provide examples of the evolution of computer memory cells to illustrate intuitively what is meant by the "structure of information." While these examples are of a digital system with discrete transitions between states, the formalism produced can also be used to represent continuous transitions such as those undergone by q-bits. Consider a 16 bit word of computer memory; that is, a $2^{16}$ state system that can honor references to its entire value, hi and lo bytes, and sixteen independent one bit values numbered from 0 to 15, and which evolves deterministically. There are many different ways that such a system can evolve. First, let us consider such a system where a program and data external to it can contribute to its evolution, i.e., an open system. As an open system there are no limits to how our system can evolve. It may, for example, successively take on the values composing an entire database. Though our system is only 16 bits, in an openly evolving system any amount of information can be received by observing the system as it changes. Open systems can be characterized in several ways. If each value of the system through time is mapped to some other state such as its previous state or the start state, then an open system is one where the mappings can be to values of other, external, systems. An open system can also be characterized as one in which the map itself has state and changes.

[1] In theory, a state-based methodology should be as expressive as a bit-based one; however, it can be shown that in many state-transition based formalisms a conflation of the concepts equality-as-structural-isomorphism and equality-as-numerical-identity causes information to be lost in the formalism. This loss of information makes it hard to model several phenomena such as intensional contexts and mereological relations.

We can make several observations from the above that will lead into a discussion of closed systems. First we notice that as a state system evolves through a series of states, new information may be required to ensure that the system remains deterministic. In the above example, a static 16 bit system contained only 16 bits, but if it changed states even once it may require up to 32 bits to remain deterministic.

**O1:** *The map, program, function, etc. that determines how a system evolves is, itself, a piece of information whose contents become inferable from the changes or patterns of change in the system in question.*

We also note that all of the information necessary to produce a sequence of values in such systems exists when the program starts. As the state of the system changes, no new information is generated. Rather, the information that already existed at the start of the program becomes expressed in the progression of state values. The exact pattern a piece of information goes through, the way it interconnects with other information in a sequence, is the *structure* of that information.

**O2:** *While we can refer to a piece of information at a particular place or time, we can also refer to that piece of information as it extends through time or space and interconnects with other pieces of information.*

Now we can develop the concept of a closed system. For a particular state system that evolves, let us include with it all the information in external systems that will have a bearing on how that system evolves; that is, all the information that determines its state. However, let us not include any extra information; that is, if a particular state of an external system could change without affecting the sequence of values our system-under-question takes, it is not included. Such systems are closed. If a system does not evolve then it is closed by itself.

It can easily be shown that if a closed system evolves, each state maps to exactly one unique state. In other words, the mapping function produces a one to one map from the list of possible states to that same list. We will use this as a definition of closed systems.

**D1:** *A* closed system *either does not evolve or evolves according to a mapping function that maps the set of possible states one-to-one onto itself.*

By using different references into a closed system we can have different views of its evolution. This will be useful in specifying information structures. Consider a closed 16-bit system. Referred to as a single entity it has $2^{16}$ states but cannot evolve. If we divide it into two bytes, we can, for example, use the hi byte as a "program" to specify something about the evolution of the lo byte. Of course there are only 256 such "programs" possible. We would then retrieve the full 16 bits by observing the state of the lo byte and observing which program it was running. We could produce longer sequences by having 15 bits for our "program" and 1 bit for the evolving state system.

This "dividing" of a piece of information into sub parts that evolve vs. programs that determine how it evolves can make the evolving piece act like an open system. That is, states can map to more than one value. We will use this as our definition of an open system.

**D2:** *If a closed system has sub-parts j and k such that k evolves based on j, then k is called an* open system*.*

From the above discussion, and from general computer science knowledge that information pieces such as bytes divide into smaller pieces such as bits, we make the following observation:

**O3:** *Pieces of information divide into sub-parts. The size of the parts can be measured in bits or in number of states, but the addition of part sizes is done by adding bits.*

Also from the preceding discussion we make this observation about describing the structure of a piece of information:

**O4:** *To completely describe the structure of a piece of information, enumerate its sub-parts and their inter-mappings.*

Suppose that two pieces of information, both of which are sub-parts of a closed system, interact. For example, suppose the hi byte of our 16-bit system affects the lo byte by copying its bit 0 into bit 0 of the lo byte. Clearly, if something is not done to preserve the lo bit 0, its old value will be lost and we will no longer have a closed system. Either that information can be simultaneously copied into the hi bit 0, or a more complex "information circuit" can be created involving three or more pieces of information. Said another way, the information which will be referenced by "lo bit 0" is the same information that was previously referenced by "hi bit 0." Some other reference must be made to refer to the old "lo bit 0." Because in this scenario we needed to talk about the "same" information we have at least an initial, intuitive reason to assert that that information has (numerical) identity. Notice that, unlike in traditional mathematics [Devlin, 1993], this (informatic) identity is not the same as equality. In fact, there are no *a priori* Identity Criteria [Guarino] in this theory. All identities will either be asserted or inferred by a process of elimination over a closed system. More than one bit may have the value 1 or the value 0, but merely because the two values are *equal* does not imply that as information pieces they are numerically *identical.*[2] In fact, suppose that both bits-to-be-swapped have the value '1.' Still, the swap would have taken place, though the numerical state of the system would not have changed. This conflation of two notions of identity can cause a loss of relevant information.

**O5:** *Pieces of information have* numerical identity*. This identity is not implied by equality of value.*

Observation O4 requires that sub-parts of information pieces be mapped to each other, but it does not specify how the mapping is to be done. The example just given suggests that specifying identities is the method of mapping. In the example, the new value of bit 0 of the lo byte is *identical* to the old value of the hi byte's bit 0. Clearly, mappings more complex than identity are needed. For example, what about the mapping that A=B+C? In the information structure formalism that follows, such complex mappings can be made by mapping sub-parts and super-parts to each other.

**O6:** *Mappings among the parts of a piece of information are made by specifying* identities *that hold among them, their sub-parts, and super-parts.*

From the observations made so far, the structure of an information system is specified by describing the identities that hold among a collection of information nodes which have sub-parts. Suppose that we have 5 nodes A, B, C, D, and E, of one byte each. Perhaps the nodes represent five "const" bytes that do not change, or perhaps they represent a single machine byte that undergoes four state changes. If we do not want to mention their sub-parts (e.g., bits), we can assert that none of them are identical, or that there are some that are identical to others, or that all are identical. In the first case, our total information is five bytes, while in the last case there is only one byte, through five copies of it. A more complex situation can occur when we refer to sub-parts. We might map the bits of A to those of B, then those of B to those of C. A simple and not very useful structure would be where there were no identities among the nodes. In a slightly more complex structure, perhaps pairs of nodes are related. For example, perhaps the bits of A map to B, those of C map to D, and perhaps the low four bits of E map to its high four bits. In more complex structures, however, longer chains or *sequences* occur. For example, the bits of A might map to B, those of B map to C which maps to D then to E. Such sequences occur in character strings storing complex data, arrays of interrelated data, streams delivering a protocol, and memory cells and other state systems as they progress through time. While such partial orderings could be represented without special features of a language, they are so common that a practical language would have facilities for representing them. In fact, isolated information pieces not identical to anything can be seen as a single element sequence, all systems that can be represented under this model can be represented as collections of interrelated sequences.

[2] An argument with a similar conclusion is made in Stonier's Information and the Internal Structure of the Universe.

**O7:** *In systems of any complexity, information is mapped into partially ordered* sequences*, and representing systems by representing sequences and their interrelations is easier than doing so by representing only information nodes.*

Consider a collection of information pieces C and a sequence S where each element in S divides into two parts S.ref and S.val where each S.val is a member of C such that whenever two S.ref's are *equal*, their corresponding S.val's are *identical*. We will call the S.ref's *references*, and the S.val's *referents*. An example of such a system is where an array is the C, instances of looking up a value in the array are the S's, and for each such instance the array index is S.ref, and the value referenced is S.val. Other examples are the mechanism for using pointers into memory and that for using a hash table.

**O8:** *There are structures where the value of one piece of information can be used to refer to another piece of information.*

## *2.2 The Theory and Formalism*

In this section, we develop a theory for representing information structures in a way that merges the features of ontology language and programming language. There are two aspects of the development: the theory apart from any notation that represents it, and the definition of a language called *Quanta* which is an embodiment of the theory. We separate the theory from its embodiment in a notation to better facilitate the creation of improved versions and new notations to be formed around the theory. While Quanta will function as a powerful ontology and modeling language, it can, at the same time be used as a full fledged programming language. Because a correctly built Quanta engine can recreate or adjust programs to fit new or special hardware situations as described in an ontology, it is suggested that Quanta would be as useful as an intermediate language in a Grid Services Platform, as well as for traditional programming tasks such as scientific and business applications or high speed virtual reality games.

In order to keep notation-independent theory separate from the Quanta specification, all discussion that is specifically about Quanta will be presented within tables. We begin by specifying a modification of BNF [Barrett] that we will use and defining syntax for Quanta tokens and comments.

The syntax for Quanta is specified in the following BNF style form:

1. Terminal symbols are underlined.
2. Non-terminal symbols are in italics.
3. Symbols in bold are being defined.
4. Alternative items are either on a new line or in the form (a | b).
5. Optional items are enclosed in square brackets.
6. Items in curly braces can appear zero or more times.
7. Case is unimportant.
8. Some items are explained in English or by example.
9. The start symbol is *infon*.

**Token**
Alpha-or-'_' { Alpha-or-'_'}

**Comments**
// Comment … | /* ….. */ ( Block comments are nestable)

### 2.2.1 Proof, Identity, and Normalization

The problem of merging ontology language with programming language is illustrated by noting the difference between algorithms that run programming interpreters or compilers and those of languages such as DAML/OWL. We wish to have a language that can be efficiently compiled or interpreted, yet the proof algorithm of DAML/OWL is a graph traversal based on modus ponens [Lissila]. Such an algorithm does not easily map to the von Neumann block/loop/conditional statement architecture of traditional programming languages. One method of creating programs in logic is the method of Prolog [Ghezzi]. However, the method used in Prolog is insufficient because, as before, without the explicit block/loop/conditional statements of imperative languages, Prolog is too hard or too inefficient to use for many purposes such as game programming, application programming, or scientific computation.

The observations that information has identity and that the structure of information is given in terms of such identities provides an alternative that will solve the problem. Rather than use a graph-traversal method based on modus ponens, we will use a graph traversal based on the substitution of identicals as in equational logic [Ramsamujh]. As we shall see, we can map the block/loop/conditional methodology to our variant of the equational logic if the items that are substituted are pieces of information and the criterion for substitutability is informatic identity. For logic and math, the equations logic works in its regular way; for example, a problem might use the identity X(Y+Z)==XY+XZ to solve an equation. Since the items that are substitutable are information we can have chains of identities mapping information through hardware. For example, Patients_info == CertainPlaceInDatabase == ResultsFromWebService == ResultsFromHttpInteraction == OperatingSystemInteraction could be a chain of identity relations that the engine uses to find a patient's information. Likewise, the identities describing how information flows through time or in through another sequence can be used to specify or control the evolution of a computer object or physical system. We summarize in the following statement:

**S1:** *All operations on information in this theory are based on the substitution of identicals in non-intensional contexts, or on operations that are compositions of such substitutions.*

Under this model, the notions of "executing code" and "evaluating expressions" are not as useful as that of "normalizing information." The code "2+3" could be "executed" by generating code that calculated the answer, or it could be "evaluated" to obtain the answer. In this theory we say that 2+3 *is* the information we want, and "5" is identical information in a normal form. The former will have the latter substituted for it. A normal form is obtained by applying identity substitutions until the desired form is reached or until there are no more identities that apply. When a system that changes over time is normalized, as will be discussed later, the resulting system is an executing program. In other words, it is not the case that Quanta programs can be compiled and then optionally run. Rather, when they are normalized they *are* executing programs whether that happened via substituting the results of interpretation, compilation, or a combination of such methods, and whether executed on a single processor on a distributed grid.

**S2:** *The decision of which identity substitutions to make is based on the goal of reaching a normal form or some other desired form for a given piece of information.*

### 2.2.2 The Quanta engine and the Context Hierarchy

In the use of a Quanta engine, a file of Quanta code is parsed and an in-memory image is created. This image can be thought of as a snapshot of the state of a Quanta engine in time. Next, that image is submitted to be normalized. When it is normalized, the results of queries will have been compiled, programs will be running or have finished, and the state of the external world may have changed or be changing. A reference to the object created will be returned which can be used to find the results of queries, get handles to programs, erase the object, or even use in future jobs. This process can be done by a person, a shell, or another program or service. Thus, a plain interface to a Quanta engine may be a CLI; however, a GUI or other type of interface can be imagined as well. In our test bed there are parse and normalize member functions. We also have a shell that calls them and a test tool that calls that shell.

A normalized Quanta object will contain names of data sets, running programs, external systems, classes, and all the identity statements defining these items. It is useful to be able to normalize new objects in the context of those already normalized in order to access their namespace. When an object is normalized, a reference to it is returned new object passed in for normalization can use that reference to assert "I am a member of this object." Since Quanta names have scope local to where they are defined, this creates a context hierarchy that can be used to run multiple objects without namespace collisions.

Consider an example situation. When the Quanta engine is first run, it loads a top level object called "World." World typically includes libraries defining basic types, local system resources, external systems, and user information. It may also contain running programs that maintain the system. As an example, a user may then create a (very simple) object such as "2+3" and submit it for normalization where it will be transformed into "5." The shell automatically dereferences the returned reference and displays the results. More complex code might include programs or commands to run, or queries that contain references to the objects modeled in World. A new reference will be returned which can be used to access the new object.

A reference to World or its members can also be used in a call to parse. Literal strings in Quanta can be delimited in a number of ways, but if a string is a serialization of a type of which World contains a map to that string format, the parser can use that map to read in the string thus making delimiters unnecessary in unambiguous cases. For example, World might contain a date type and a map to a string format. Then, in code parsed under the context of World, if the parser can tell that a certain string is a date, dates in that form can be used as literals and quotes are not needed around the string.

**Q1:** *A Quanta engine parses Quanta code into a memory image of a snapshot of a system at an instant. It normalizes such snapshots to create dynamic models and returns a reference to the model. Given such a reference, parsing and normalizing can be done in the context of (i.e., as a member of) an already-normalized object to access its namespace. Various UI types can be made to provide tools for parsing, normalizing, accessing, and deleting objects.*

The exact method used by a Quanta engine to choose and apply identity substitutions is not prescribed here, and is implementation dependent. However, literal forms should generally be preferred to symbolic forms.

### 2.2.3 Shortcuts and Connection to the Outside World

Often, given a list of assertions in the current informatics theory, statements can be inferred that are not necessarily in the given list. This is analogous to theorems in mathematics.

**D3:** *Statements that hold given a collection of assertions in this (informatics) theory are called* shortcuts.

Because information processing devices can access actual information from the world outside the processing engine, references in the logic can refer to or manipulate objects or systems in the external world. For example, a name in the language might produce information from an I/O stream when dereferenced. Information from those systems can be recognized as such by asserting or inferring an identity relation between the information from the dereferenced name and information from the external entity. For example, information from such a name might be asserted or inferred to come from a video camera, and the information entering the video camera might be asserted or inferred to come from some physical scene. Perhaps, given a sufficient model, the contents of the images can be inferred from the stream. This is an advantage over traditional logics where a barrier between language and reality exists [Read, 5-34].

**D4:** *Names of information sources that automatically produce information from or push side-effects onto the external world are called* autonames.

### 2.2.4 Quanta Shortcuts, Autonames, and Auto-Optimization

The Quanta syntax can be used to describe complex information structures by implying various identity relations. In addition to these relations, there are a number other sources of identity assertion that the engine can use to optimize performance. They are *shortcuts* and Autonames.

Shortcuts are assertions kept by an engine that could be deduced from given identities. They are akin to theorems in mathematics. For example, given the axioms defining arithmetic, the Quanta shortcut:

```
for x:<numbers> :: %x * 0 == 0;
```

could be deduced. It states the theorem x * 0 == 0. Other shortcuts might provide functions for such tasks as retrieving web pages via http, calling specific web services, or constructing application objects.

Autonames are Quanta names that are recognized by the engine and are given special identities. For example, in our test bed for Quanta, the calculation of sum of two numbers is evaluated without recourse to the axioms of arithmetic using a fast addition function written in C++ (for addition of literals). Autonames are one way for an engine to provide a connection to a file or I/O system.

In advanced engines, logs can be kept of what identities are applied for various problems. These log files could be examined to find series of identity substitutions that are applied often, and a shortcut could be created automatically. Some such shortcuts may be links to binary code. Likewise, for names that are dereferenced often, a dereferencing procedure of compiled into binary code could be made into an Autoname.

### 2.2.5 Infons

A term is needed that is less ambiguous than either "piece of information" or "state system." For example, state system could mean a state-system-at-an-instant, a state-system-over-its-lifetime, a time-slice from a state-system, a state-system-over-an-interval etc. Furthermore, the term "state system" connotes a physical system separate from any content that its state may hold. Keith Devlin uses the term *infon* to refer to "a piece of information." [Devlin: 1991] That term will be used here. More precisely, we use it to mean a collection of one or more constant states or pieces of information. By "constant", we mean unchanging or even a-temporal. Thus, an infon can be a state system at an instant or over an interval in which the same piece of information is held. However, it can also be a system over time in which each item in the infon is one of the states of the system as it changes over time. The notion of time is not built into the concept of an infon; we will have to add it later. The items in an infon need not be connected with each other in any way. They could be the states of a system over time (time-slices), but they could as easily be the eight bits of a byte or the bytes in a megabyte. The usage for infons will be analogous to that of sets. In fact, the theory is modeled after ZF set theory [Devlin: 1993], but where all objects are infons and literal values are used instead of the empty set to generate the theory[3]. Sub-

[3] ZF set theory uses the empty set as the foundation of all its objects. For practical reasons, the current theory uses literal values for that

sequently infons will be defined more formally; however, here the definition of infon is informal. Intuitively, one can think of an infon as a state system, but perhaps the best translation is "piece of information." Since almost everything can be modeled as a state system [Malitz], almost everything is considered an infon: computer memory, chunks of memory, data-structures, objects, web-services, even people, atoms, and institutions.

**D5:** *Infons are pieces of information. That is, they are state systems viewed from the bit-centric point of view.*

Many equivalent methods could be used to describe the structure of infons. The following lists the infons of Quanta.

**Q3: Infons**

*Syntax*

**Infon**
( *Infon* ) // Infons can be enclosed in parenthesis.
*CollectionInfon*
*DifferenceInfon*
*IntersectionInfon*
*ComplementInfon*
*BoolInfon*
*LargeIntInfon*
*StringInfon*
*NameInfon*
*IdentityInfon*
*ForInfon*
*VarInfon*
*MatchInfon*
*ArrayInfon*
*ContainInfon*
*PartialIDInfon*
*ConditionalInfon*
*SequenceClassInfon*
*SequenceContainsInfon*
*IntRange*
*BoolQryInfon*
*CommandInfon*

*Identities implied*

1. All identities implied by the infon type chosen are implied for this infon.

### 2.2.6 Sub-Infons and Identity

By observation O3, infons can have sub-infons just as state-systems can have sub-systems, i.e., information can be subdivided into smaller information pieces such as where objects have members and bytes divide into bits. It should be emphasized that an infon *just is* its members. It is not a structure such as a list which would take some structural memory even if the list contained nothing. An infon with no members contains zero bits and has only one state. This mereological type of containment also implies that infon membership cannot be circular, though infons *can* contain circular *references*.

We will use words such as containment, membership, is an element of, and so on to signify sub-infonship.

**S3:** *Infons can have sub-infons which are also infons. (O3)*

The relation between an infon and its members is:

**S4:** *Two infons are informatically identical if and only if they have the (informatically identical) same sub-infons.*

---

purpose. It would be interesting to remake this theory using (zero-bit) null infons {} as the foundation.

S4 allows us to infer the members of an infon if we know the members of an identical infon. We can refer to the identities implied through containment with either of the notations:

Contains: *some specification*

and

Is Member of: *some specification*.

### 2.2.7 Literals

Literal values are used as infons with which the engine has direct contact. At least some literals must have an equal operator defined. Quanta provides literals for indefinitely large integers, strings, and Boolean values.

**S5:** *Literal values are infons.*

**Q4: Literals**

*Syntax*

**BoolInfon**
( true | false )

**LargeIntInfon**
A string of digits terminated by a non-digit.

**StringInfon**
*Token*
"This is a string\n" // All C escapes are supported.
'This is a string\n' // Strings can contain embedded new-lines.
A *here document* as in UNIX scripts.
'Chunked' binary stream as in HTTP.
Automatic parsing of context-defined, serializable types.

*Identities implied*

1. LargeInts have arithmetic operations defined for them, as well as gt, lt, ge, and le comparison operators. Later in this paper the algebraic identities will be defined. Other functions may be added to the language later.
2. Strings are "sequences" which are defined later. This implies a range of access functions and identities providing access to characters, sub-strings, and even named sections defined by a sort of regular expression.

*Autonames*

1. The arithmetic and comparison operators for LargeInts are implemented as autonames.

### 2.2.8 Reference and Quanta Names

By observation O8, we can use one infon to refer to another infon. In logic and traditional set theory, entities can be named at will. In an actual physical system, however, references are themselves information and take up memory in a state system.

**S5:** *Infons can refer to other infons according to the principle that all* equal *names are defined to refer to* identical *infons.*

This notion of reference is important for several reasons. First, it can be used to spell out the "logic" of references so that the engine can make inferences about them. Secondly, information cannot divide into parts indefinitely but must stop at bits.

One requirement is to be able to refer to an infon and then refer to its sub-infons. If logical simplicity were the main disiratum it would suffice to have the ability to declare that a particular infon is to be used as a reference. That reference could save sub-parts such that one part selected the parent infon while the next part selected the desired sub-part. It is significantly more efficient, however, to build such hierarchical referencing into the language. In Quanta, names are segmented using the 'dot' as a separator as is typical in programming languages such as C++. Successive segments refer to members of the infons referred to be the preceding segments. Here is the notation for Quanta's references, called *names*:

### Q5: Names

*Syntax*

**NameInfon**
<*NameSegment*>

**NameSegment**
*Infon* [$] [. *NameSegment*]

*Identities implied*

1. All *equal* names refer to *identical* infons by definition; thus, they can substitute for each other. Names are equal if they have the same number of segments and corresponding segments are equal, perhaps determined by the equal operators for literals.
2. If a name <N> refers to an infon j, then adding a segment <O> to <N> ensures that <N.O> refers to a sub-infon of j. That is, <N> always contains <N.O>.

*Comments and Examples*

1. Names in Quanta have the same logical properties as names and functions in programming languages. If they refer to changes, they can even act like functions that affect the state of their parent or functions with external side-effects. Logically they are similar to maps in set theory.
2. Quanta names are not abstract entities, but entities that exist as information structures in a state system.
3. Some examples suggest usage:

   <World.England.Queen>
   <Sqrt.9>
   <Sum.[5,3]>

   (The square brackets in these examples delimit Quanta's literal arrays, which are defined later.)
4. The infons of a name segment can be of any type. So, names can be used as arrays, and maps.
5. In cases where the syntax is unambiguous, alternate syntaxes are acceptable and will be used in this paper. Examples are:

   Sqrt(9)
   Sum(5.3)
   5+3
   World.England.Queen
6. The optional dollar sign ($) in name segments will be explained in Q9.

### 2.2.9 Membership and Identity

Because equality does not imply identity, equality operators cannot directly be used to determine identity. Therefore, it is necessary to explicitly declare identity much of the time. Declaring identity between two items means that they have the same (identical) members (S4). It is useful to be able to also declare that one item contains all the members of the other item, *and perhaps more* (partial identity). Likewise, it is useful to be able to declare that the one contains the other and may also contain other items (containment).

Statement S4 generates three binary operators asserting identity, partial identity, and containment. Naturally, these three operators are bidirectional; if A is identical to B, then B is identical to A, if A is partially identical to B then the members of B are members of A, and if A contains B then B is a member of A. Because of this, there are two different ways inferences that can be made for each operator. While these inferences should not be ruled out, we can make the engine's task easier by preferring one direction of inference to the other. There are three ways that a language can benefit from preferring a single direction of inference for identity and containment. First, an engine must navigate a maze of identities in order to arrive at a normal or desired from for a given piece of information. This is essentially a graph traversal where each identity is an edge. By preferring a single direction through each edge, and by verifying that each edge leads to an actual transform, it can be shown that, because the goal is usually a normal form of the information, and each transform, presumably, bring the infon closer to that form, an edge will not usually be applicable more than once. Except for the special cases, this relieves the engine of having to keep track of every node it visited in the graph. For example, a transform that maps "2+3" to "5" will not be encountered again in the same context because "5" does not match "2+3." The special case involves transforms involve the order of elements. For example, the assertion that addition is commutative, which in Quanta reads:

for x:<numbers>, y:<numbers>:: %x + %y == %y + %x

would apply to 2+3 as well as to 3+2, thus the engine would be in an endless loop. This problem can be overcome by detecting order-based assertions and only applying those only if the parameters are unsorted by some ordering criteria. Thus, 3+2 would be transformed into 2+3, but since [2, 3] is sorted, the transform would not occur again.

The second benefit to having a preferred direction of inference for identity and containment operators is that it provides an automatic method for specifying normal forms. Programmers simply arrange assertions so that the transform most often desired is applied. For example, since the form "5" is preferred to "2+3" in most cases, the assertion of identity would be arranged to that substituting the former for the latter happened as the preference. When alternate forms for information are desired, the identity can be transversed in the opposite direction.

Lastly, having a preferred direction of inference allows the engine to guess when names are being defined vs. used. This is especially useful when defining sequences. In sequences, items of the sequence are often mapped to their predecessor. It would be unwieldy to always have to specify when a "current" value vs. a "preceding" value was being referred to. With a preferred direction of inference, it can be assumed that one side of the identity statement refers to the "new" value, and the other side the "old" value. Thus the identity assertion "<A> == <A>+1" can mean close to what it means in traditional programming. The "l-value" gets the normalization of the "r-value."

### Q6: Identity Assertions

*Syntax*

// L-Value and R-Value are synonyms for infon.

**IdentityInfon**
*L-Value* == *R-Value*

*Identities implied*

1. L-Value and R-Value are identical infons.

*Comments and Examples*

1. Though identity is symmetric, transforms in Quanta proceed from L-Value to R-Value in normal cases.
2. Later, sequences and “sequential contexts” will be discussed. When an L-Value is a name in a sequential context, then it refers to the “next” value in the sequence. R-Values in a sequential context refer to the “current” value in the sequence.
3. Some simple examples are:

   <Company.CEO> == ”Fred Smith”
   <Person.Age>==<Person.Age>+1
4. This example assumes the reader knows about collections, which are defined later; however, it is an intuitive example.

   <Person>=={<name>==”Bob Smith”; <age>==42; <ShoeSize>==9;}

   If the above assertion is true, then “<Person.age>==42” is true. This defines three fields for <Person>. There may be more fields, but they will map to these three. That is, these three are sufficient for determining identity.

### Q7: Partial Identity

*Syntax*

// L-Value and R-Value are synonyms for infon.

**PartialIdInfon**
*L-Value* := *R-Value*

*Identities implied*

1. L-Value contains all the members of R-Value, and perhaps more.

*Comments and Examples*

1. Transforms relying on partial identities in Quanta proceed from L-Value to R-Value in normal cases. That is, the preferred inference here is that L-Value contains something, not that members of R-Value are members of L-Value.
2. When an L-Value is a name in a sequential context, then it refers to the “next” value in the sequence. R-Values in a sequential context refer to the “current” value in the sequence.
3. An example similar to that for identity assertions.

   <Person>:={<name>==”Bob Smith”; <age>==42; <ShoeSize>==9;}

   Unlike with the identity assertion, the possibility that there are more fields that do not map to these three is left open here. The identities of name, age, and ShoeSize are not enough, by themselves, to determine the identity of Person.

### Q8: Containment

*Syntax*

// Type-Class and Item are synonyms for infon.

**ContainInfon**
*Type-Class* : [*Item*]

*Identities implied*

1. Item is a member of Type-Class.

*Comments and Examples*

1. The preferred inference for containment is that item is a member of Type-Class so that anything the holds for all members of Type-Class holds for Item. This is used to generalize.
2. Statistical inferences are possible. Suppose 50% of the items in Type-Class are green and the rest are blue. Then asserting that Item is a Type-Class asserts those probabilities over Item. The current Quanta engine does not support such inferences.
3. Some examples are:
   <Customers> : <Bob>; // Bob is a customer.
   <Dog> : <Fido>;
   <Loves> : {<Lover>==<Alice>; <Loved>==<Bob>}; // A relation.
4. The “item” can be left out if a name is not needed.

### 2.2.10 Sequences of Infons as Partial Orderings

Observation O7 in section 1 implies that because infons map to each other, often in long sequences, the ability to represent partially ordered sequences is important in an information structure language. An example of such usage is in representing sequences of events through time as described in Lamport’s *Time, Clocks, and the Ordering of Events in a Distributed System* [Lamport]. Lamport represents such an ordering using a “Happened-Before” relation between events. Sequences of infons are not necessarily temporal, and therefore are not necessarily separated by events. For example, sequences of characters in a string are not, in general, temporally situated. Later we shall add the notion of events to sequences to help represent temporality. Currently, we shall adapt Lamport’s partial ordering to collections of infons by associating a *next* and *previous* field with each infon in a sequence. We also associate a *first* and *last* field with each sequence, along with *size* and a numerical reference to each infon in the sequence. Lastly, we define that there is a method for referring to sub-sequences within a given sequence.

**S6:** *A collection C of infons is a sequence S if each infon in C is associated a next and previous infon in C and S has references to first and last infons in C such that first.previous and last.next are {}, and the other next and previous fields work to form a doubly linked list which includes every member of C. S is also associated with a numeric size field which is the number of infons in the sequence.*

**S7:** *Each infon in a sequence S can be referenced by a numeric index in the standard way where S[0]==first, and S[size-1]==last.*

**S8:** *A sub-collection of the items in a sequence S is a sub-sequence if all of its members are sequential in the doubly linked list. Sub-sequences can be referred to by a start and end reference into S or by a start index and a size specification for the sub-sequence.*

It is important to realize that what is being defined is a collection of names and the relations among their referents, not a fancy doubly linked list. The names will be used as a model. Perhaps no values in the model will be known, but perhaps some will. For example, if it asserted that a particular sequence has size=100, then the system knows only that fact. A query for the fifth item, for example, will fail. Likewise, if the system knows all the values in a sequence but not, explicitly, the size field, it may be able to deduce the size by counting the items in the collection. The mode of operation is to gather as much information as is needed, and to infer other infons by applying the principle of substitution of identity. Suppose it is asserted that a certain field contains the size of a sequence, and that a different area of memory contains the items in the sequence. A query for the sequence might trigger a query for the size which would be substituted for the results of reading the size field. Similarly, the first part of a sequence may be stored in one database while the second part of the sequence is stored in, for example, a local file. These identity mappings can be asserted, allowing the engine to find the data for querying or manipulation.

In section 2, a number of constructs for representing information structures are developed. For many of these constructs, there is a simple way of creating a partial ordering out of it. For example, a notation for collecting pieces of information into a single group can be modified to also *order* those pieces. For consistency, our syntax will be to add a dollar sign ($) to a construct when we desire to activate the extended semantics. For example, a string "Hello" can have extra semantics applied to it by $"Hello", and a collection {A, B, C} can be ordered by ${A, B, C}. One can think of the $ as short for "sequence."

Now we suggest an exception to the normal method of naming entities that may seem strange at first. We have defined a number of fields that sequences have such that (in normal Quanta) we would expect to refer to an entire sequence S with the notation <S>. Likewise, we would expect to refer to the first node (for example) in S with the notation <S.first>. The problem is that in many cases we wish to refer to a particular item in the sequence, and we wish to reference it, not by a numeric index or by a series of ".next" fields, but as the infon "just after" the previously referred to infon. Therefore, for many information structure languages it will be useful to allow the normal reference to a sequence refer to this "current element" and use a special notation to refer to the structure as a whole. As is specified in Q9, we use a $ to indicate this difference. We now consider the Quanta notation for representing sequences.

### Q9: Sequence Structure

*Syntax notes*

1. When describing or populating a sequence (vs. using one), it is useful to be able to refer to the infon "just after" the previously referred to infon. Thus, for names of sequences a special case is made in the way their names refer to them. If a name segment that refers to a sequence ends with a dollar sign ($), the reference is to the entire sequence. Otherwise, the reference is to a particular element in the sequence. The determination of this element will be discussed in Q13. This syntax was defined in Q5.
2. A number of syntactic constructs for generating sequences will be considered after this. However, all of them have in common that they produce a number of named sub-infons, or fields. More precisely, a sequence contains the following sub-infons.

*Identities implied*

For a sequence S:

1. <S$.first> == A node for the first infon in the sequence.
2. <S$.last> == A node for the last infon in the sequence.
3. <S$.size> == The number of infons in the sequence.

If *n* and *m* are number from 0 to size-1:

4. <S$.*n* > == A node for the *n*th element of the sequence.
5. <S$.[*n, m*] > == A subsequence starting at position n and of size m.

For a sequence node N:

6. <N.next> == A node for the next infon in the sequence or {}.

7. <N.prev> == A node for the previous infon in the sequence or {}.
8. <N.value> == The actual infon in the sequence.

---

### 2.2.11 Declaring the Type of the Elements in a Sequence

Infons, in our notation, are static entities that do not change. They might be considered "const" entities. If we wish to model an entity that changes over time we shall have to represent it as a sequence of infons. Suppose, for example, that we wish to represent an integer stored in computer memory that can change over time. In C++ the notation to declare such an entity is:

```
int i;            // i is an integer that can change.
```

In Quanta, we can create a class <ints> which is the infon containing all integers. We could then assert:

```
<ints> : <i>;     // <i> is a const integer.
```

While this looks similar to the C++ notation, it declares that <i> is a single integer value, that is, a "const" integer. The equivalent information structure is a sequence of integer infons. One way to declare such a thing is to declare a sequence then assert that all of its elements are integers. This method is tedious, and we would like for Quanta to look as similar to traditional programming languages as possible, so we develop two shorthand notations for Quanta.

### Q10: Sequence Types

*Syntax*

// Type-Class and Item are synonyms for infon.

**SequenceClassInfon**
*$NameInfon*

**SequenceContainsInfon**
*Type-Class* <u>**$:**</u> *Item*

*Identities implied*

1. The resulting SequenceClassInfon is a sequence of whatever type is named by NameInfon. That is, all the members of the resulting sequence are members of the infon named by NameInfon.
2. In the resulting SequenceContainsInfon, Item is a sequence of which, all members are of type Type-Class.

*Comments and Examples*

1. Because $<ints> is the class of all sequences-where-all-members-are-integers, a SequenceClassInfon can be used to declare a sequence of integers.

   ```
   $<ints> : <i>;    // i is an integer that changes value over time.
   ```

2. An even easier method is to use a SequenceContainsInfon which is similar to the ContainsInfon, but which automatically declares a sequence.

   ```
   <ints> $: <i>;    // i is an integer that changes value over time.
   ```

### 2.2.12 Integer Ranges and Sequences

Collections and sequences of integers are quite common, and Quanta has a special notation for declaring them.

### Q11: Integer Ranges and Sequences

*Syntax*

// StartInt and EndInt are synonyms for infons evaluating to integers.

**IntRange**
**[$]** *StartInt* **..** *EndInt*

*Identities implied*

1. The resulting infon is collection of integers in the specified range.
2. If a dollar sign ($) is included, the collection is also a sequence.

*Comments and Examples*

1. 5..10 // The collection of integers {5, 6, 7, 8, 9, 10}
2. $5..10 // A sequence of integers from 5 to 10.
3. $10..5 // A sequence of integers from 10 to 5.

#### 2.2.13 Blocks: Collections + Set Theoretic Union

Observation O3, which states that infons divide into sub-infons, suggests that as long as we do so in a mereological fashion we can form new infons by collecting other infons.

**S9:** *Each infon has a specific number of "bits" associated with it, and accounting for identities among its sub-infons, this number is identical to the number of bits of its sub-infons. (That is, bits must add correctly.)*

**S10:** *For a collection C of infons an infon k can be formed which has as its members all of the members of C (provided that S9 is not violated.)*

By S10 we can list individual infons to create a new infon, but O3 also suggests that we can list a group of infons by specifying their "parent" infon.

**S11:** *A collection of infons C can be specified by a list of members and/or by referring to a "parent" infon P and asserting that all the members of P are in the collection C.*

In Quanta we use the familiar notation of enclosing items in curly braces. If we mark an item with a "#" symbol it means that we include not the item itself, but all of its sub-infons (in accordance with S11).

Observation O7 suggests that many collections C of infons have identity relations such that the sub-infons of C form sequences of infons, and that language constructs for specifying relations among infons and their sub-infons can be usefully enhanced by adding semantics for asserting a partial order among the sub-infons. In Quanta we specify that a collection (or block) is a sequence by prepending a "$" to it, as in ${Eenie, Menie, Minie, Moe}. Now we look at the block grouping construct.

### Q12: Blocks, Collections, and Infon Union

*Syntax*

**CollectionInfon**
**[$] { { [#]** *Infon* **;} [ ...; ] }**
**[$] { [#]** *Infon* **{ , [#]** *Infon* **} [, ...] }**

*Identities implied*

1. The infon resulting from a CollectionInfon contains all listed infons that are not marked with a "#", and all the *members* of the listed infons that *are* marked with a "#".
2. If the last element in a CollectionInfon list is the token "…", then there may be other elements in the collection which are not present. If the element is not present, then the list is exhaustive with regard to non-identical elements, and the whole can be considered identical to the parts.
3. If the resulting CollectionInfon C is marked with a "$" sign, then C is a sequence where the order of the items in the list corresponds to the order in C. If C is marked with a "$" sign, then all items in C marked with a "#" must also be sequences. Such items are subsequences in C.
4. In the next section, it is defined that certain sequences S1 can contain other sequences S2. Because the S2 sequences must be contained in a sequence, a CollectionInfon containing such a sequence will automatically be a sequence, even if the dollar sign marker is not present.

*Comments and Examples*

1. There are two syntaxes for CollectionInfons. The first uses semicolons (;) as terminating symbols while the second uses commas (,) as separating symbols. The choice of syntax has no semantic affect, and is a matter of preference. Generally, the semicolon syntax looks better for long statements:

   { <a>==5; <b>==6; <c>==7;}

   And the comma syntax looks better for short lists:

   { red, green, blue}

2. As an example of identity statement 1, in

   <A>=={red, green, blue}

   <A> contains three elements, which are the strings 'red', 'green', and 'blue'. Likewise, in:

   <B>=={red, green, {a, b, c}, blue}

   <B> contains four elements, which are the strings 'red', 'green', and 'blue', *and* the collection {a, b, c}. However, in:

   <C>=={red, green, #{a, b, c}, blue}

   <C> contains *six* elements, which are the strings 'red', 'green', 'a', 'b', 'c', and 'blue'.

3. As defined in identity statement 2, while

   <A>=={<X>==red; <Y>==green; <Z>==blue;}

   implies that <A> is identical to the three elements such that <A.X>==red, <A.Y>==green, and <A.Z>==blue, and no other infons exist that are not identical to one of these three. However, in:

   <A>=={<X>==red; <Y>==green; <Z>==blue; …;}

   while <A.X>==red, and so on, the "…" implies that there may be other infons as well such that the inference that <A> is identical to the three items cannot be made.

4. Identity statement 3 implies that

<S>==${red, green, blue}

is a sequence where <S$.first>==red, etc. Further:

${red, green, ${a, b, c}, blue} == ${red, green, a, b, c, blue}

### 2.2.14 Automatic References and Inter-Sequential Relations

While a sequence is simply an ordered collection, there are many uses for them when attempting to model complex information structures. For example, a string of characters can be a sequence. The characters in a string need not have any sub-structure, but they usually do. What would be useful is to be able to name the relevant sub-sequences and elements in a string in a useful way, then to map those names together as appropriate. We can easily provide Quanta names for the parts of a string by using sequences in a way analogous to regular expressions. Thus, for example, one section of the string might consist of a decimal number which is to be used as the length of another section to come later. Or perhaps a section of a string contains elements to be mapped into a spreadsheet (e.g., a string mapped to a spreadsheet file.) Names of arrays, lists, trees, and so on can be declared and mapped into the string. Quanta takes advantage of the fact that the basic elements of a regular expression are sequence, repetition, and disjunction (or conditional) to map Quanta blocks, Quanta for-loop style structures, and Quanta conditionals to reproduce the functionality of regular expressions. (The for structures and conditionals will be defined later.)

Many times, mapping a sequence to itself through regular expression defined names is insufficient. It is usually the case that sequences map to other variables, and even other sequences. In addition to mapping strings or other sequences to themselves, another use for mapping the elements in a sequence is mapping the sequence of values that a system goes through as it evolves through time. For example, an integer in memory evolves through a sequence of values. It is especially true that such "temporal" sequences often map, not to themselves, but to each other and to constant values. If we can successfully notate the relations among all the sequences (temporal or not) and constants we will be able to represent any information structure.

As with regular expressions, the Quanta block, for-structures, and conditionals provide the necessary logic. To obtain all these features we do not need to add any extra semantics to these constructs. The general method is to use one sequence P, which contains the sequences to be mapped C1, C2, … Cn, such that the order of P gives order to successive additional statements about the Cx's. In other words, P contains regular expressions for each Cx, but the segments of the regular expressions are interspersed so that they can refer to each others parts. The formal description with examples will clarify this. However, because of the nature of the representation the notation is analogous to Von Neumann-style "block, loop, conditional" architecture, and is thus quite intuitive. Still, it is important to remember that we are not specifying an algorithm, but a collection of names and associated identity statements which we can use as a model of an information structure. The model could be used to assert "an entity like this exists", but as easily it could be used to create or query such entities.

## Q13: Ordered References to Parts of Sequences

*Syntax*

No new syntax is required, this applies to all definitions of sequences. Namely, CollectionInfons, ForInfons, and strings.

*Identities implied*

1. If the definition of a sequence S contains references to S's sub-infons which are also sequences T, named <T>, the following rules apply.
   a. References to T which are the entire left side of an IdentityInfon or a PartialIDInfon are labeled L-values. All other references to T are labeled R-values.
   b. If <T> is an R-value, it refers to the item in the sequence T that was last referred to as an L-Value.
   c. If <T$> is an R-value, it refers to the entire sequence T.
   d. If <T> is an L-value, it refers to the item in the sequence T that is the .next field of the element in T last referred to as an L-Value.
   e. If <T$> is an L-value, it refers to a sub-sequence whose first member is the .next field of the element in T last referred to as an L-Value, and which has the same size as the sequence on the right side of the an IdentityInfon or PartialIDInfon.
   f. The sequence S is defined to contain all of any such member-sequences so that the last reference to a member-sequence marks the end of the member-sequence.
2. As mentioned in Q12, a CollectionInfon or ForInfon which contains the type of construct just described is automatically a sequence even if the "$" mark is left off.

*Comments and Examples*

1. In the following example, the entire sequence <FunSequence> is initialized at once.

```
{
   <strings> $: <FunSequence$> ==
            {Eenie, Meenie, Minie, Moe};
}
```

2. This example defines the same sequence in parts.

```
{
    <strings> $: <FunSequence> == Eenie;
    <FunSequence> == Menie;
    <FunSequence$> ==  {Minie, Moe};
}
```

3. Here, values from several sequences are mapped in various ways. The connection with traditional assignment is apparent.

```
{
  <ints> $: <Bobs_Age>==1;
  <Bobs_Age>==<Bobs_Age>+1;

  <ints> $: <Sams_Age>==<Bobs_Age>
}
```

After several more constructs are developed, extended examples of regular expression style mapping and multiple-sequence structures are given.

### 2.2.15 Strings and Sequences

Strings in Quanta are sequences. Thus, for example, given:

<PersonsName> == "Ralph";

the following are true:
<PersonsName$.size>==5;
<PersonsName$.first>==R;
<PersonsName$.[1,3]>==alp;

Quanta has a special syntax for generating strings as sequences.

**Q14: Strings as Sequences**

*Syntax*

**String (extended syntax)**
**$** *StringInfon*

StringInfon can contain Quanta names or Quanta code enclosed in percent signs (%).

*Identities implied*

1. If the embedded code can be transformed into a string or has an <asString> member the string will be inserted as a substring into the original string.
2. Embedded code can produce new strings that also have embedded code which will be evaluated. Thus, entire protocols can be produced with just a string and names that expand it.

*Comments and Examples*

1. $"The sum of 5 and 3 is % 5+3 %."
2. 
```
{
  <FirstName>==Bruce;
  <LastName>==Long;
  <FullName>==$"%<FirstName>% %<LastName>%";
  <Greeting>== $"Hello %<FullName>%. Welcome!";
}
```

### 2.2.16 For Infons and Variables

In addition to references and the collection/membership/identity constructs, the "for/variable" construct is one of the primary infon constructors. It is analogous to the universal quantifier from logic, as well as to the for loop from traditional programming languages. Given one or more infons to use as the range for variables, the for construct declares an infon for each permutation of the variables' members. All of those infons together make up the forInfon. The forInfon can also be made into a sequence by using ranges for variables that are sequences.

**Q15: For Infons and Variables**

*Syntax*

**VarDef**
**[$]** *token* **:** *infon*

**ForInfon**
**[$]** for *VarDef* {**,** *VarDef*} **::** *Infon*

Applies the variables defined in VarDefs to the Infon which may contain VarInfons. Each variable ranges over all the members of the infon associated with it.

```
for x:<numbers>, y:<numbers> ::
    <sum.[%x,%y]>==<sum.[%y,%x]>
```

**VarInfon**
**%** *token*

*Identities implied*

1. a forInfon contains in instance of its *infon* for each permutation of the variables in its header. If those infons contain varInfons, then each variInfon is replaced with the value of the variable whose token it matches.
2. In addition to the identities in the forInfon's *infon* field, any identities that hold among the elements of the variable ranges hold among the new infons.
3. If a forInfon is marked by a "$", then that forInfon is a sequence. The order of the sequence is determined by which tokens in the variable list are also marked by "$". Going across the variable list from left to right, each permutation of the "$" marked variables produces a sub-sequence in the forInfon. Successive variables are nested in their predecessor's sub-sequence.

*Comments and Examples*

1. This example produces 5 integer infons, all with the value 10:

   for x:1..5 :: 10;

   Note: the "1..5" is an IntRange infon.
2. The following infon has an infinity of members and is part of a definition of mathematical groups and therefore of numbers:

   ```
   for x:<numbers>, y:<numbers> ::
       <sum.[%x,%y]>==<sum.[%y,%x]>
   ```
3. Here we define a "square" function:

   ```
   for x:<numbers> ::
      <square.%x> == %x * %x
   ```
4. Here we set <A> to be a sequence that ranges, in order, over a two dimensional array where the x dimension of the array contains <x_size> elements, and the y dimension contains 5 elements:

   ```
   <A> ==
       $for $x : 0..(<x_size>-1),
           $y : 0..4 ::
               <MyData.[%x, %y]>
   ```
   Here, <A$.first> would be identical to <MyData.[0,0]>

### 2.2.17 Data Queries

A query for information in Quanta is simply an infon that is normalized and its resultant form is fetched for viewing or processing. A simple example is:

5+3

A more complex query will rely on the existence of a model:

<OurCompany.Schedule.<dates.["May 12, 2004"]>>

Or

<CDrive.Mydocs.SprdSheet1.cells.[E,32]>

Given an appropriate model, queries could be made to fetch any type of information from sources such as P2P networks, web or grid services, databases, etc.

To fetch an entire dataset, a predefined collection (infon) can be requested or a "for structure" can be created to join and process data. As we shall show, algorithms can be specified so that data can be further processed in a query.

However, Quanta is not merely a model-and-query language. Complex programs such as applications or 3D games could be described and instantiated.

### 2.2.18 Boolean Queries

In many queries, and in particular in the condition part of a Conditional infon, a Boolean value is needed that tells whether a particular situation obtains. For example, one may want to query whether the Evening Star is identical to the Morning Star:

<OurSystem.EveningStar>==<OurSystem.MorningStar>.

One could merely submit the query and if the resulting reference to an infon was null, it would mean the system thus described did not exist and thus it was false. However, it is often preferable to have results in a true/false form. To represent such a query in Quanta, simply append a question mark ("?") to the end of any infon descriptor.

<OurSystem.EveningStar>==<OurSystem.MorningStar>?.

The construct could be placed in an if-else infon or used in any other query.

Notice that the shift from retrieving information and verifying that the retrieval was successful to a retrieving of either true of false causes a loss of information. Suppose that a quantity <n> is unknown. The query:

5+3

will normalize to "8", but the query

5+<n>

is already in normal form because the engine cannot normalize it further. Thus, we can tell from the result that the query was not completed because <n> could not be found. When we shift to a pure true/false model, such information is lost.

**Q16: Boolean Query Infons**

*Syntax*

**BooleanQryInfon**
*Infon* ?

*Identities implied*

1. The normal form of a BooleanQryInfon is a BoolInfon. If the result is "true", then *infon* exists as it is described. If the result is "false" the *infon* does not exist as described.

*Comments and Examples*

1. The fact that many BooleanQryInfons result in true and many in false is a good example of why identity cannot be inferred from equality. If it could, then "2+2?", which evaluates to true would mean the same as "<NuclearReactions>:<sun>?", which would imply that the question "what is 2+2?" has the same content as "is the sun a nuclear reaction?" This is an example of why state-based representations (vs. bit-based representations) struggle to represent the meaning of content.

### 2.2.19 Conditionals

The logical meaning of conditionals used here is only indirectly related to the truth-functional, disjunctive meaning from the predicate calculus [Jeffrey]. Logically, the boolInfon is used as an index into an array of classes, and the chosen class is asserted.

In practice, the engine can merely evaluate the condition then consult the correct block infon. The syntax should seem familiar.

**Q17: Conditional Infons**

*Syntax*

**ConditionalInfon**

[$] If *BoolInfon*, *infon* [ else *infon* ]

*Identities implied*

1. If BoolInfon is true, this is identical to the first infon. If it is false, if there is an else clause, it is identical to the infon in the else clause. Otherwise, see rule 2.
2. If neither the first or second infons can be applied, then if the if clause is marked by a "$" the result is an empty sequence ${}. Otherwise the result is {}. If the dollar sign is present, both infons should evaluate to sequences.

### 2.2.20 The All Matching Infon

As with ZF set theory, a method of selecting infons that match a certain description is needed [Devlin, 1993]. Classes of objects can be made by defining "the collection of all infons where X obtains" then declaring an infon to be a member of that class.

**Q18: All Matching Infons**

*Syntax*

**MatchInfon**
@ *Infon*

*Identities implied*

1. Contains all and only those infons of which *infon* is a description.

*Comments and Examples*

1. The following defines <RedThings> as a class that contains all and only infon with a color field that is Red. Next a particular apple is asserted to be in that class.

```
{
    <RedThings> = @{<colors>:<color>==Red};
    <RedThings>:<MyApple>;
}
```

### 2.2.21 Referencing Infons

If a name for an infon appears by itself it is ambiguous whether it is being defined or used. Consider the infon:

{<MyInfon>}

This could be interpreted as a definition for <MyInfon> as a member of the main object, or it could be a reference to a name already defined elsewhere. In Quanta, it is the latter. In order to specify that <MyInfon> is a sub-name in an object (as in the former case), simply use it in an identity statement:

{ <MyInfon>=={…} }.

If an identity statement is not given, the implied meaning is that the infon being named can be substituted for this name. In the following example,

```
<Top> == {
    <MyInfon>==Hello;
    <OurInfon>=={<MyInfon>};
  }
```

if we were to look up the value of <Top.OurInfon> it would be

{Hello}.

### 2.2.22 Assertions from Classes

Similar to set theory, the mechanism for generalizing in Quanta is to define an infon that contains all and only the infons that have the property to be generalized over, then declare that the item to be described is a member of that infon. This method is used for all generalization whether generalizing static types such as ints or structs, generalizing sequences of actions into a function, describing classes of objects, or even templates over classes or "templates of templates" and so on.

In most cases, a class such as <ints>, for example, which is defined as a class that contains all 32 bit integers, is used by asserting that it contains a particular infon. For example, to declare an integer called <ShoeSize>, use the form:

<ints> : <ShoeSize>

This notation can be combined with an assertion of identity:

<ints> : <ShowSize> == 12

In some cases, however, an anonymous variable is needed. For such a case, the name can be left out entirely:

<ints> :

As we will see shortly, this notation is used to call functions that cause any object to change state. For example:

<print.["Hello World!"]>:

Or with an alternate syntax:

print("Hello World!"):

Here, <print.["Hello World!"]> is a collection of all occasions where "Hello World" is printed in a particular context. (Obviously, this collection cannot be completely dereferenced.) The added colon means something like "One of these occurrences goes here." Until this makes sense, the intuition that such a construct is a function call will suffice.

### 2.2.23 Set Theoretic Constructions

In addition to the union operation provided by CollectionInfons, it is useful to have Difference, Intersection, and Complement constructors.

**Q19: Set Theoretic Constructions**

*Syntax*

**DifferenceInfon**
difference ( *Infon* , *Infon* )

**IntersectionInfon**
intersection *Infon*

**ComplementInfon**
( complement | not ) *Infon*

*Identities implied*

1. A DifferenceInfon contains all and only those members of the left infon that are not in the right infon.
2. An IntersectionInfon contains all and only those infons that are members of every member of Infon.
3. A ComplementInfon contains all and only those infons that are *not* members of Infon

### 2.2.24 Arrays

While sequences provide a partial order, and even a full order if one considers that numeric indices can usually be used to map into a sequence, it is useful to have a simple method of creating simple, zero-based arrays. Such literal arrays have been used several times already to provide parameter lists for Quanta names. For example, in the name:

<sum.[5,3]>

the "[5,3]" is a literal array. The advantage of using arrays to pass parameters is that a non-literal array can be passed in to represent very large or equation-defined lists of parameters. For example, an infinite series could be summed by using a for structure that ranged over the integers to specify an infinite array of terms to add. An array is simply an infon whose members have numeric indices.

**Q20: Literal Arrays**

*Syntax*

**ArrayInfon**
[ [*Infon*] {, *Infon*} ]

*Identities implied*

1. Each infon in the list is assigned a numeric name starting from zero. There are no sub-sequences allowed in the notation. In fact, for efficient processing, arrays are not sequences.

*Comments and Examples*

1. The following defines a simple array

   [red, green, blue]

   This is equivalent to

   {<0>==red; <1>==green;, <2>==blue}.

   So, for example, if <A>==[red, green, blue]; then <A.0>==red.

### 2.2.25 Commands: Affecting the World

Thus far we have used identities to model, query for information, and to query for Boolean "true or false" information. Lastly, we look at using identities to do the reverse of querying, commanding. Just as the actions of other computers and physical systems can be modeled as an information structure, the actions of the engine running the query can also be modeled as information structures. So given a description of an information structure, perhaps in Quanta, it makes sense to assert "make this description true" even though such an assertion does not immediately seem like an assertion. Because there is no major type distinction between these two kinds of information structure, the engine can use the same algorithm of substituting identicals to process such "queries."

We add a new infon type to Quanta called a *command infon.* It is merely an infon marked with a "!".

**Q21: Command Infons**

*Syntax*

**CommandInfon**
*Infon* !

Precedence for '!" is lower than for all other infons.

*Identities implied*

1. This type of infon is identical to the event it describes, and therefore it is identical to all infons that would describe the same event. This is why when such an infon is normalized, the event "happens." (The actual working of this identity is simple and intuitive, but the linguistic expression of it seems strange. The examples below will clarify.)

*Comments and Examples*

1. The following example is of simple assignment:

   ```
   <A>==5!
   ```

   Suppose an infon exists that maps the above infon to a “set” function:

   ```
   for x:<ints>:: (<A>==%x!) == <Set.[A.%x>]>;
   ```

   Then the infon <A>==5! will be transformed into the form of a call to the set function. Such functions may be autonames that modify memory, call an operating system, or perhaps call a C function. Or, perhaps they are not autonames, but they call autonames.

2. The following example might be read “allocate a const integer named <A>:

   ```
   <ints> : <A> !
   ```

   While the following means allocate a non-const integer:

   ```
   <ints> $: <A> !
   ```

   The next example means allocate a sequence of integers:

   ```
   <ints> $: <A$> !
   ```

   The precise mechanism for such allocations is implementation dependent; however, complex allocations should perhaps be built out of identity assertions rather than hard coded. This will allow the engine to decide the optimal execution.

3. Commands made from collections, for-infons, and the other infons will also be handled by substituting identities or by autonames.

4. When named classes are used to make an assertion is in described in 2.2.22, such classes which contain commands can be thought of as function calls with side-effects. Programmers need not describe all the identities such a “function call” makes. For example, a function that writes a line of text to a consol might look like:

   ```
   <writeln.[“Hello World”]>:
   ```

   Or in the alternate syntax,

   ```
   writeln(“Hello World”):
   ```

5. Classes called as specified in 2.2.22 which do *not* have commands in them can, nevertheless, alter the state of the world if they are called as commands by appending a “!”. For example, a class that describes a number begin incremented can be commanded:

   ```
   increment (<A>):!
   ```

## 3 Techniques for using Quanta

We now move from a theoretical mode to a practical one where we look at how common programming tasks can be implemented in Quanta.

### *3.1 Describing Functions without Side Effects*

To describe a function that does not change the state of any system, merely use a for-infon to define names that range over all the values the function can take as its inputs. While functions of any complexity can be defined, we illustrate using a simple example. Here is a function definition that takes an integer parameter and normalizes to its value plus one.

```
for x:<ints> ::
    <oneMore. %x> == %x +1;
```

### *3.2 Describing Sequences for Serializing and Parsing*

Describing a sequence involve describing part or all of its syntax and part or all of its contents. To describe syntax we need to be able to assert concatenation, repeated structures, conditional structures, and nested structures. To describe the contents, the sections of the sequence described by the syntax specification can be named and mapped to literal values, other items in the sequence, functions, etc. They can also be declared to be a member of a class in order to limit their values to some sub-set of the values allowed for their type. Such a mapping can be used to both when reading and when writing sequences.

The example that follows illustrates using Quanta collections to assert concatenation, and Quanta for-infons to assert repetition. We define a sequence where the first item is an integer named Length and the successive items are chars numbered from 0 to Length.

```
${
    <ints>:<Length>,
    #$for $i: $0..<Length> :: <char>:<%i>
 }
```

Next, we use the” Match –All” infon (@) to create a class of all items like those just described, and we name it <charArrays>.

```
<charArrays> == @  // <charArrays> is the infon of all infons where:
   ${
        <ints>:<Length>,
        #$for $i: $0..<Length> :: <char>:<%i>
     }
```

Now let us use the sequence class we just defined to describe a new sequence. Our new sequence will contain an identifying header that is the string “HEAD”, then it will contain a single character, and lastly, if the character was “A” then the following item is a <charArray> otherwise nothing. We will name our sequence <Seq>.

```
<Seq$> == ${
    <strings>:<Header> == “HEAD”;
    <char>:<condition>;
    #$if (<condition> == “A” )?
        <charArray>;
 }
```

Let us now provide more detail be asserting
<Seq$> == ${HEAD, A, 5, #${a,b,c,d,e}}

In the sequence we just described, the following is true:

```
{
   <Seq.Header> == <Seq.0> == <Seq.first> == “HEAD”;
   <Seq.condition> == <Seq.1> == <Seq.first.next> == A;
   <Seq.3> == b;
}?
```

The sequence construction tools in these examples can be used to describe any describable sequence. By creating multiple sequences, naming their parts, and mapping the names to each other, multiple sequences can be interrelated. However, in cases where the parts of sequences are not known in advance, because perhaps the sequences contain conditionals or items of varying length, such mappings can be awkward. The constructs described in Q13 make this task easier. We examine these next.

### *3.3 Describing Algorithmic Structures*

The memory cells of a computer, as well as any physical system that changes state over time can be seen as sequences ordered through time. When a number of such systems interact through time, the system comprised of all of them can be described by documenting the interrelations among their respective sequences. Recall that the notation described in Q13 implies that the two following sequence notations are equivalent:

```
<chars> $: <Seq$> == ${A, B, C, D}
```

and

```
${
    <chars>$:<Seq> == A;
    <Seq> == B;
    <Seq> == C;
    <Seq> == D;
  }
```

However, in the latter notation, sequences can be interspersed. For example, the two sequences:

```
<chars> $: <Start> == ${A, B, C,}
```

and

```
<chars> $: <End> == ${ X, Y, Z}
```

can be combined using the notation:

```
${
    <chars>$:<Start> == A;
    <chars>$:<End> == X;
    <Start> == B;
    <End> == Y;
    <Start> == C;
    <End> == Z;
  }
```

This interspersing of elements allows us to describe interrelated sequences in way that resembles the traditional von Neumann method for specifying algorithms. In the following example, note how blocks, with local sequences (variables) can be nested, and how for-structures and conditional infons nest and operate in the expected way given the identities they imply as specified in part 2.

```
{
    <strings>:<CourseIDs> == {math, english, science, history};

    #$for $Course : <CourseIDs> ::
     {
         <ints> $: <RunningTotal> == 0;
         #$for $Student : <Courses.%Course.Students> ::
         {
              #$if (<Courses.%Course.%Student.Status> == completed)?
                   <RunningTotal> == <RunningTotal> +
                             <Courses.%Course.%Student.Score>
         }
         <Courses.%Course.TotalScores> == <RunningTotal>;
     }
}
```

This use of syntax seems even more familiar when commands are used in the “algorithms”:

```
${
  // print a simple multiplication table
    #$for $x:$0..10 :: {
         #$for $y:$0..10 ::
              write(%x * %y, “ “):;
         writeln():;
    }
}
```

While the above item certainly appears to be an algorithm, it is better characterized as a description of an information structure. This is because the engine is free to manipulate the system by substituting identicals. Thus, for example, a particular structure may be distributed over a large computing grid, or the items may be done in a different, but equivalent, order. Furthermore, the above is an assertion of fact; it could be made into a query (“?”) or what would normally be expected, a command (“!”).

### *3.4 Mapping to the Outside World*

Suppose an object exists on another machine, or perhaps in the non-virtual world, and we which to read and alter the state of that system. Realistic examples are an online database, a robot body, or a remote computer. All of these examples are merely complex state systems, and control of them comes in the form of operations that alter the state in a predictable way, or read the state in some way. We will consider the simple case of reading and altering a byte of memory.

First, we declare that the byte exists. We will name it <B>:

```
<byte> $: <B>;
```

This asserts that a sequence ($:) of bytes exists, and names it <B>. If we wanted to assert that <B> was a constant, we could have asserted:

```
<byte> : <B>;
```

We can assert that our sequence starts (as of the time of the assertion) with a particular value:

```
<byte> $: <B> == 5;
```

This does not *make* <B> == 5, rather, it informs the engine of the fact. The assertion could potentially be wrong.

We can assert that the next (second) value is one more than the previous:

```
<B> == <B>+1;
```

Again, this merely alerts the engine to a fact. If we wish to tell the engine to *make* the next infon in the sequence be a certain value, we use the command marker (“!”):

```
<B> == 10!;
```

But how does the engine know how to read or write such values? The notation <B>==10! means roughly, “create a new infon, which *is* the next item in the sequence <B$> and make it identical to 10.” Thus, this infon can be mapped as well as any other. We posit a function (perhaps an autoname) called WebSvc that calls a web service.

```
#for x:0..255 :: {
   (<B> == %x!) ==<WebSvc.[ChangeValue,<IpAddr>, …, %x]>;
   <B> == <WebSvc.[ReadValue, <IpAddr>, …]>;
  }
```

The first line in the *for infon* maps changes to <B> to a web service, while the second line maps reads of <B> to the service.

This example illustrates mapping a simple assignment to an external object. The engine could do a lot with such a map by iterating over many such calls. However, by mapping more complex Quanta structures such as for infons or collections, entire datasets can be manipulated at once. Furthermore, such maps need not call functions, but could create a small compilation unit which, when combined with others could be linked into a binary that executes complex tasks, and the executed. Similarly, actions can be mapped to other machines, and more than one mapping can be defined so that a cleverly designed engine can choose the optimal ones. Thus, a task can be defined apart from any machine that may run it, and the engine can choose the optimal execution method whether that is function calls, compilation on the local machine, or farming out all the sub-parts to a Grid. In this way, Quanta programs can be made machine independent, and yet be made to take advantage of special hardware so that the potential exists for the creation of faster-than-C execution.

If a particular system cannot be altered, the engineer coding in Quanta should not provide a map for it. Thus, for example, a function which will send an email to Aristotle should not be written, as its assertions would be false (unless the engineer also invents a time machine!)

If an engine is capable of allocating memory to create objects as commanded by:

```
    <ints> $: <B>!
```

the engine must also provide an implicit or explicit mapping to be used to read and change its value.

### *3.5 Describing Functions with Side Effects*

A function that modifies a system can be written by creating an algorithm that calls functions that are mapped to external systems. Suppose we wish to have <B> as defined in the previous section take on a series of values:

```
$for $v: 1..10 ::
    <B>==%v!;
```

If we wish to generalize this action we must create the class of all such actions and declare a member of that class to exist at the right time.

```
// <CountB> is the action.

<CountB> == @
    $for $v: 1..10 ::
        <B>==%v!;

// We now call that action two times:
<CountB>:;
<CountB>:;
```

Suppose we wish to add a parameter to the function so that the sequence need not end at 10.

```
for maxVal : 1..255 ::
    <CountB.%maxVal> == @
        $for $v: 1..%maxVal ::
            <B>==%v!;
```

Functions where the internal algorithm is not modeled can also be created by mapping them to autonames. This technique could be used to create functions such as “print” or “beep” without having to specify their information structure.

### *3.6 Describing, Creating, and Manipulating Objects*

An object traditionally has the properties of polymorphism, encapsulation, and inheritance. Inheritance relations can be defined using the set theoretic operations. Polymorphism can be achieved by carefully defining the types and constraints used to define names of sub-infons for objects. Lastly, encapsulation can be achieved by writing functions that modify the object and making those function sub-infons of the object. By asserting that every items in the sequences of member variables is identical to one of the “member functions”, it will be a contradiction to directly modify a variable.

We create a class of objects called <SimpleObjs> which contain two integers, one of which has an “increment” member function.

```
<SimpleObjs> == @
{
    <ints> $: <NumOfPets>;
    <ints> $: <age>;

    <incAge> == @((<age>==<age>+1) ! );

}
```

We create, initialize, and manipulate a <SimpleObjs>:

```
{
    <SimpleObjs> : MySO == {<NumOfPets>==2; <age>==35;};
    <MySO.incAge>:;
    print (<MySO.age>);

}
```

### *3.7 Other Techniques*

There are a number of techniques that are not fully developed here. For example, it would be useful to represent intensional contexts such as beliefs, desires, intentions, etc. It would also be useful to be able to represent systems with continuous state transitions such as qubits.